\documentclass[useAMS,usenatbib]{mnras}

\usepackage{graphicx}	
\usepackage{amsmath}	
\usepackage{amssymb}	
\usepackage[dvipsnames]{xcolor}

\newcommand{\be}{\begin{equation}}
\newcommand{\ee}{\end{equation}}
\newcommand{\bea}{\begin{eqnarray}}
\newcommand{\eea}{\end{eqnarray}}

\def\aap{A\&A}
\def\apj{ApJ}

\def\apjl{ApJ}
\def\mnras{MNRAS}

\def\apjs{ApJS}
\def\prd{Phys. Rev. D}

\def\ltsima{$\; \buildrel < \over \sim \;$}
\def\simlt{\lower.5ex\hbox{\ltsima}}
\def\gtsima{$\; \buildrel > \over \sim \;$}
\def\simgt{\lower.5ex\hbox{\gtsima}}

\title[Cosmic Degeneracies II: structure formation in $f(R)$-WDM]{Cosmic Degeneracies II: Structure formation in joint simulations of Warm Dark Matter and $f(R)$ gravity}
\author[M. Baldi \& F. Villaescusa-Navarro]{\parbox{\textwidth}{Marco Baldi$^{1,2,3}$, Francisco Villaescusa-Navarro$^{4,5}$}
\\
\\$^{1}$Dipartimento di Fisica e Astronomia, Alma Mater Studiorum Universit\`a di Bologna, viale Berti Pichat, 6/2, I-40127 Bologna, Italy;
\\$^{2}$INAF - Osservatorio Astronomico di Bologna, via Ranzani 1, I-40127 Bologna, Italy;
\\$^{3}$INFN - Sezione di Bologna, viale Berti Pichat 6/2, I-40127 Bologna, Italy;
\\$^{4}$INAF - Osservatorio Astronomico di Trieste, Via Tiepolo 11, I-34143, Trieste, Italy;
\\$^{5}$INFN - Sezione di Trieste, Via Valerio 2, I-34127 Trieste, Italy;}

\begin{document}
\pagerange{\pageref{firstpage}--\pageref{lastpage}} \pubyear{2011}
\maketitle
\label{firstpage}
\begin{abstract}	

\noindent We present for the first time the outcomes of a cosmological N-body simulation that simultaneously implements a Warm Dark Matter (WDM) particle candidate and a modified gravitational interaction in the form of $f(R)$ gravity, and compare its results with the individual effects of these two independent extensions of the standard $\Lambda $CDM scenario, and with the reference cosmology itself. 
We consider a rather extreme value of the WDM particle mass ($m_{\rm WDM}=0.4$ keV) and a single realisation of $f(R)$ gravity with $|\bar{f}_{R0}|=10^{-5}$, and we investigate the impact of these models and of their combination on a wide range of cosmological observables with the aim to identify possible observational degeneracies. In particular, we focus on the large-scale matter distribution, as well as on the statistical and structural properties of collapsed halos and cosmic voids. Differently from the case of combining $f(R)$ gravity with massive neutrinos -- previously investigated in \citet{Baldi_etal_2014} -- we find that most of the considered observables do not show any significant degeneracy  due to the fact that WDM and $f(R)$ gravity are characterised by individual observational footprints with a very different functional dependence on cosmic scales and halo masses. In particular, this is the case for the nonlinear matter power spectrum in real space, for the halo and sub-halo mass functions, for the halo density profiles and for the concentration-mass relation. However, other observables -- like e.g. the halo bias -- do show some level of degeneracy between the two models, while a very strong degeneracy is observed for the nonlinear matter power spectrum in redshift space, for the density profiles of small cosmic voids -- with radius below $\approx 5$ Mpc$/h$ -- and for the voids abundance as a function of the void core density. 
\end{abstract}



\section{Introduction}
\label{i}

The presently accepted standard cosmological scenario -- known as $\Lambda$CDM -- has been proved as an extremely successful model capable to explain numerous cosmological observations, from the statistical properties of the anisotropies in the Cosmic Microwave Background \citep[CMB,][]{Planck_2015_XIII} to the clustering properties of galaxies or cosmic neutral hydrogen on large-scales \citep[][]{Alam_etal_2016,Delubac_etal_2015}. This model assumes that the observed accelerated expansion of the Universe \citep[][]{Riess_etal_1998,Perlmutter_etal_1999,Schmidt_etal_1998} is driven by a cosmological constant $\Lambda$ whose value is set by observations, and that the dark matter that drives the growth of cosmic structures is {\em cold}, i.e. that it has negligible thermal velocities at all redshifts.

Despite the spectacular success of the $\Lambda$CDM model, some of its predictions are in tension with observations on both large and small scales. In particular, a persisting tension has recently emerged between the best-fit cosmological parameters obtained for a $\Lambda $CDM cosmology from CMB observations \citep[][]{Planck_2015_XIII} and a number of large-scale structure observations including weak gravitational lensing \citep[][]{Heymans_etal_2013,Hildebrandt_etal_2016}, redshift-space distortions induced by the peculiar motion of galaxies \citep[][]{Blake_etal_2011b,Reid_etal_2012,Simpson_etal_2016} or galaxy clusters \citep[][]{Vikhlinin_etal_2009}, as well as from galaxy cluster counts \citep[][]{Planck_XXIV}. Furthermore a number of longstanding observational tensions at small scales are still persisting.  Among these there are: {\em i}) the cusp-core problem: N-body simulations predict that the density profile of dark matter halos exhibits a cusp in their cores, while from observations we know that some galaxies present a core in their density profile \citep[][]{Salucci_etal_2007,Gilmore_etal_2007,Eymeren_etal_2009,Naray_etal_2010,Walker_etal_2011}. {\em ii}) The missing satellite problem: the number of satellites around Milky-way dark matter halos from simulations is much larger than the number we observe \citep[][]{Moore_etal_1999,Klypin_etal_1999}. {\em iii}) The too-big-to-fail problem: the kinematic properties of the most massive subhalos around Milky-way halos from simulations are in strong tension with observations \citep[][]{BoylanKolchin_Bullock_Kaplinghat_2011,BoylanKolchin_Bullock_Kaplinghat_2012}. 

In order to overcome these tensions and to alleviate the theoretical naturalness problems associated with the observed energy scale of the cosmological constant, several modifications of the standard $\Lambda $CDM cosmology have been proposed and  investigated over the past decades. In particular, in the present work we will be focusing on Warm Dark Matter (WDM)  scenarios aiming to solve the small-scale issues of the standard model, and on Modified Gravity (MG) theories as possible alternative and more natural explanations for the observed accelerated cosmic expansion, also possibly providing new interpretations to the above mentioned large-scale tensions. 

Both these alternative cosmological scenarios have been widely investigated in the literature and their effects on a large number of observables have been  tested and clarified. In particular, WDM models have been tested by numerous authors \citep[see e.g.][and references therein]{Colin_AvilaReese_Valenzuela_2000,Bode_Ostriker_Turok_2001,Avila_Reese_etal_2001,Yoshida_etal_2003,Viel_etal_2012,Schneider_etal_2012,Lovell_etal_2012,Maccio_etal_2012,Maio_Viel_2015,Carucci_etal_2015} and their impact on the formation of structures at small scales has been compared with observations \citep[][]{Narayanan_etal_2000,Viel_etal_2005,Miranda_Maccio_2007,Markovic_etal_2011,Viel_etal_2013,Kennedy_etal_2014} thereby placing constraints on the viable WDM particle mass. Similarly, in more recent years MG cosmologies have attracted significant interest for their expected impact on structure formation and on the evolution of collapsed structures. In particular, the $f(R)$ gravity theory that will be discussed in the present work has been implemented in simulation codes of structure formation by several authors \citep[][]{Ecosmog,Puchwein_Baldi_Springel_2013,Llinares_Mota_Winther_2014} and a wide range of simulated observables within these MG scenarios have been obtained \citep[just to metion some, see e.g.][]{Li_Zhao_Koyama_2011,Li_etal_2012,Jennings_etal_2012,Fontanot_etal_2013,Arnold_Puchwein_Springel_2014,Cai_etal_2014,Hellwing_etal_2014,Arnold_Puchwein_Springel_2015,Cai_Padilla_Li_2015,Achitouv_etal_2016}.

In the present work we aim to investigate the joint effects of these two independent modifications of the standard cosmological model, testing for possible observational degeneracies and quantifying the deviations that each model produces on the expected signatures of the other. This type of analysis has already been performed for the combination of $f(R)$ gravity theories with a non-negligible value of the total neutrino mass \citep[][]{Baldi_etal_2014} allowing to identify a very strong degeneracy between these two classes of models. We will therefore proceed along the lines of \citet{Baldi_etal_2014} and explore the joint effects of $f(R)$ gravity and WDM particle candidates using high-resolution cosmological simulations.

The paper is organised as follows. In Section~\ref{sec:models} we will provide a brief overview on the two physical models considered in this work, namely $f(R)$ gravity in Section~\ref{sec:fR} and WDM in Section~\ref{sec:WDM}. In Section~\ref{sec:sims} we will describe the numerical setup adopted in the present work and the approaches employed for the identification of collapsed halos and cosmic voids. In Section~\ref{sec:results} we will illustrate the outcomes of our analysis on a wide range of cosmological observables. Finally, in Section~\ref{sec:conclusions} we will summarise our results and drive our conclusions.

\section{The cosmological models}
\label{sec:models}

\subsection{$f(R)$ gravity}
\label{sec:fR}

For what concerns possible modifications of the theory of gravity, as anticipated above, we will consider extensions to standard General Relativity (GR)
in the form of $f(R)$ gravity, which represents the most widely studied class of Modified Gravity models also down to their impact on linear \citep[][]{Pogosian_Silvestri_2008,Hu_Raveri_Silvestri_2016} and non-linear structure formation \citep[see e.g.][]{Oyaizu_etal_2008,Schmidt_etal_2009,Li_etal_2012,Puchwein_Baldi_Springel_2013,Llinares_Mota_Winther_2014}. 

$f(R)$ gravity is characterised by 
the action
\begin{equation}
\label{fRaction}
  S = \int {\rm d}^4x \, \sqrt{-g} \left( \frac{R+f(R)}{16 \pi G} + {\cal L}_m \right),
\end{equation}
where the standard Einstein-Hilbert term $R$ (with $R$ being the Ricci scalar curvature) is
replaced by $R+f(R)$. In Eq.~(\ref{fRaction}), $G$ is Newton's gravitational constant, $g$ is the determinant of the metric
tensor $g_{\mu \nu }$, and ${\cal L}_m$ is the
Lagrangian density of all matter fields. 
The model can be described by an additional scalar degree of freedom associated with the quantity $f_R \equiv {\rm d}f(R)/{\rm d}R$. In the weak-field and quasi-static limit this scalar field obeys an independent dynamic equation\footnote{We work in units where the speed of light is set to unity, $c=1$.}
{\citep[see again][]{Hu_Sawicki_2007}:}
\begin{equation}
  \nabla^2 f_R = \frac{1}{3}\left(\delta R - 8 \pi G \delta \rho \right) \,,
\label{eq:fR_field_eq}
\end{equation}
where $\delta R$ and $\delta \rho$ are the relative perturbations in the scalar curvature and
matter density, respectively.

A popular choice within all possible forms of the function $f(R)$ was proposed by \citet{Hu_Sawicki_2007}:
\begin{equation}
\label{fRHS}
f(R) = -m^2 \frac{c_1 \left(\frac{R}{m^2}\right)^n}{c_2 \left(\frac{R}{m^2}\right)^n + 1},
\end{equation}
where $ m^2 \equiv H_0^2 \Omega _{\rm M}$ is a mass scale while $c_{1}$, $c_{2}$, {and $n$} are non-negative constant free parameters of the model. The choice of Eq.~(\ref{fRHS}) has the appealing feature of allowing to recover with arbitrary precision the expansion history of a $\Lambda $CDM cosmology by choosing $c_{1}/c_{2} = 6\Omega _{\Lambda }/\Omega _{\rm M}$ under the condition $c_2 (R/m^2)^n \gg 1$, so that the scalar field $f_{R}$ takes the approximate form:
\begin{equation}
  f_R \approx -n \frac{c_1}{c_2^2}\left(\frac{m^2}{R}\right)^{n+1}.
\label{eq:fR-R,n_relation}
\end{equation}

In this work we will only consider models with $n=1$, which leaves $c_{2}$ as the only free parameter of the model. The latter can be also expressed in terms of the associated value of the mean scalar degree of freedom at the present epoch, $\bar{f}_{R0}$. As this convention has become the standard one in studies of $f(R)$ gravity models, we will also specify our cosmologies by their $\bar{f}_{R0}$ value.

In $f(R)$ models, the dynamical gravitational potential $\Phi $
satisfies \citep{Hu_Sawicki_2007}:
\begin{equation}
  \nabla^2 \Phi = \frac{16 \pi G}{3} \delta \rho - \frac{1}{6}\delta R \,,
\label{eq:phi_poisson_eq}
\end{equation}
so that the total gravitational force is governed by a modified potential $\Phi = \Phi _{\rm N} -\delta R/6$ (while the lensing potential $\Psi $ remains unchanged).

For our simulations we will employ the {\small MG-GADGET} code \citep[][]{Puchwein_Baldi_Springel_2013} that consistently includes the effects of the modified potential and its associated {\em Chameleon} screening mechanism \citep[][]{Khoury_Weltman_2004}. The {\small MG-GADGET} code features a  Newton-Gauss-Seidl iterative scheme to solve Eq.~(\ref{eq:fR_field_eq}) for a generic density field produced by a set of discrete particles, and computes the total force experienced by each particle through Eq.~(\ref{eq:phi_poisson_eq}) by including in the gravitational source term the curvature perturbation $\delta R$  derived according to Eq.~(\ref{eq:fR_field_eq}).
We refer the interested reader to the {\small MG-GADGET} code paper for a more thorough presentation of the numerical implementation.

\begin{table}
\begin{center}
\begin{tabular}{cc}
\hline
Parameter & Value\\
\hline
$H_{0}$ & 67.1 km s$^{-1}$ Mpc$^{-1}$\\
$\Omega _{\rm M} $ & 0.3175 \\
$\Omega _{\rm DE} $ & 0.6825 \\
$ \Omega _{b} $ &0.049 \\
\hline
${\cal A}_{s}$ & $2.215 \times 10^{-9}$\\
$n_{s}$ & 0.966\\
\hline
\end{tabular}
\end{center}
\caption{The set of cosmological parameters adopted in the present work, consistent with the latest results of the Planck collaboration \citep[][]{Planck_016}. { Here $n_{s}$ is the spectral index of primordial density perturbations while ${\cal A}_{s}$ is the amplitude of scalar perturbations at the redshift of the CMB.}}
\label{tab:parameters}
\end{table}

\subsection{Warm Dark Matter}
\label{sec:WDM}

A possible way to solve both the cusp-core and missing satellite problems, while preserving the success of CDM on large scales, is to enable the possibility that dark matter has non-negligible thermal velocities. This is what is referred to by the term Warm Dark Matter (WDM). In WDM models, the formation of halos/subhalos on scales smaller than the dark matter free-streaming length will be extremely suppressed, since only processes as halo fragmentation may produce such objects.  On the other hand, if dark matter has a non-zero temperature (i.e. thermal velocities are non negligible), the dark matter phase-space is finite, which implies that the density profile of halos can not be cuspy, but it should exhibit an inner core. Thus, both the missing satellite and the cusp-core problems would be naturally alleviated by invoking non-negligible dark matter thermal velocities.

In terms of matter power spectrum, models with WDM will exhibit a cut-off on small scales, produced by the dark matter thermal velocities, that inhibits the matter clustering on scales smaller than the free-streaming length. 
The cut-off in the primordial density power spectrum can be expressed as a transfer function with the form \citep[][]{Bode_Ostriker_Turok_2001}:
\begin{eqnarray}
T^2_{\rm lin}(k)\equiv P_{\rm WDM}(k)/P_{\rm \Lambda CDM}(k)=(1+(\alpha\,k)^{2\nu })^{-10/\nu}
\nonumber , \\  
\nonumber \\
\alpha(m_{\rm WDM})=0.048\,\left(\frac{1
  \rm{keV}}{m_{\rm WDM}}\right)^{1.15}\,\left(\frac{\Omega_{\rm WDM}}{0.4}\right)^{0.15}\left(\frac{h}{0.65}\right)^{1.3}\nonumber \\
\
\label{BOT01}
\end{eqnarray}
with $\nu =1.2$. Other forms of the linear power suppression have been found in the literature \citep[see e.g.][]{Hansen_etal_2002} with a similar qualitative behaviour.

This signature can be used to put constraints on the mass (or magnitude of the thermal velocities) of the WDM particles. The tightest constraints to date arise from measurements of the amplitude and shape of the Ly$\alpha$ power spectrum, resulting in a $2\sigma $ limit of $m_{\rm WDM}\ge3.3$ keV \citep[][]{Viel_etal_2013}. Constraints from weak lensing \citep[][]{Inoue_etal_2015} and high-redshift galaxy counts \citep[][]{Schultz_etal_2014}, although less tight ($m_{\rm WDM} > 1.3$ keV), are also consistent with this bound. 

We emphasize that the above constraints have been derived for $\Lambda$WDM models, where CDM is simply replaced by WDM. The purpose of this paper is to investigate the statistical properties of matter and halos, subhalos, and voids, in cosmologies with both modified gravity and WDM, for which the above quoted limits may no longer apply.

\section{The simulations}
\label{sec:sims}

We perform four high-resolution simulations for four different cosmological models characterised by different laws of gravity and dark matter particle mass. 

As a reference model, we consider a standard $\Lambda $CDM scenario, i.e. a model where gravity is governed by standard General Relativity and the dark matter particle mass is formally infinite, in the sense that thermal velocities are negligible and the linear matter power spectrum does not exhibit any cut-off on small scales. 

For the same dark matter particle mass we investigate also the case where gravity is described by the $f(R)$ theory described above in Section~\ref{sec:fR} for a scalar amplitude at $z=0$ of $|\bar{f}_{R0}|=10^{-5}$. 

In addition, we will consider two cosmological models characterised by these two different theories of gravity (GR and $|\bar{f}_{R0}|=10^{-5}$) and by a dark matter particle mass of $0.4$ keV, thereby corresponding to a WDM candidate with a ``very warm" temperature. Although being strongly disfavoured by the Lyman-alpha, weak lensing, and number counts observational constraints already mentioned above \citep[][]{Viel_etal_2013,Inoue_etal_2015,Schultz_etal_2014}, we consider such a low value of the dark matter particle mass in order to more easily highlight its effects on the various observables we will consider in the present work, as the thermal cut-off for such low mass is already significant at scales that are more easily resolved by our simulations. Therefore, our setup will provide a very prominent example of how the degeneracy with the underlying theory of gravity might affect some observational features that are usually employed to derive constraints on the dark matter particle mass. Furthermore, this value corresponds to the most extreme value considered in the recent work by \citet{Yang_etal_2015} about the impact of WDM on the properties of cosmic voids to which we aim comparing some of our results.

Our simulations evolve a set of $512^{3}$ particles in a periodic cosmological box of $100$ comoving Mpc$/h$ aside, with a mass resolution of $m \approx 6.3 \times 10^{8}$ M$_{\odot }/h$ and a gravitational softening of $\epsilon _{g} = 6$ kpc$/h$.
All simulations share the same cosmological parameters \citep[summarised in Table~\ref{tab:parameters}, consistent with Planck cosmological constraints, ][]{Planck_2015_XIII} and the same background expansion history since both WDM and $f(R)$ gravity (for the \citeauthor{Hu_Sawicki_2007} setup adopted in the present work) do not appreciably affect the background evolution of the universe. 

For the WDM simulations we imprint the thermal cut-off described by Eq.~(\ref{BOT01}) onto the primordial power spectrum when generating the initial conditions, which are produced by displacing particles from a regular cartesian grid according to Zel'dovich approximation \citep[][]{Zeldovich_1970} at a starting redshift of $z=127$. On top of the peculiar velocities, we add to the particles thermal velocities whose magnitude is randomly drawn from the WDM momentum distribution. The orientation of the thermal velocity vector is taken randomly within the sphere.

{It is a well known fact that standard N-body simulations of WDM suffer of numerical artifacts such as the formation of spurious halos along the filaments of the cosmic web \citep[][]{Wang_White_2007}. Such artificial fragmentation then results in a population of small halos that do not correspond to real physical objects. Numerical approaches to overcome this problem have been discussed e.g. by \citet{Angulo_Hahn_Abel_2013}. {We do not employ such alternative approaches as we are mostly interested in the degeneracy with the underlying theory of gravity and a possible increase of the abundance of low mass halos due to artificial fragmentation would simply make our results more conservative.}}

Due to limitations of computational resources and the highly demanding integration of the combined simulation we stop our runs at $z=0.25$ and we will perform all our analysis at this redshift. This is not expected to affect significantly the main conclusions of our investigation.

\subsection{The halo catalogues}

For all our four simulations we identify particle groups through a Friends-of-Friends algorithm \citep[][]{Davis_etal_1985} with a linking length of $0.2$ times the mean inter-particle separation, and store groups composed by at least 32 particles. On top of these catalogues we run the {\small SUBFIND} algorithm \citep[][]{Springel_etal_2001} to identify gravitationally bound substrucures of each FoF group, and store halos with at least 20 particles for which we compute spherical overdensity quantities as e.g. the virial mass $M_{200}$ and virial radius $R_{200}$ referred to the critical density of the universe, defined by the relation:
\begin{equation}
\frac{4\pi }{3}R_{200}^{3}\Delta \rho_{\rm crit} = M_{200}
\end{equation}
with $\Delta = 200$ and $\rho_{\rm crit} = 3H^{2}/(8\pi G)$.

For our analysis we will only use halos whose main substructure has a virial mass $M_{200}$ in the range $\left[ 10^{11}\,, 10^{14}\right] $ M$_{\odot }/h$: for higher halo masses the number of halos is too low to provide robust statistical conclusions while lower masses are poorly resolved. With such catalogues at hand we will investigate the effects of WDM and of its combination with $f(R)$ gravity on a number of statistical and structural properties of the halo populations arising in the various cosmologies, as described below in Section~\ref{sec:halos}

\subsection{Voids identification and selection}
\label{sec:voids}

We identify cosmic voids in our simulations suite by means of the 
publicly available void finder {\small VIDE}
\citep[][]{VIDE} by running it on a random subsample of the dark matter particle distribution
extracted from the $z=0.25$ snapshot of the various runs.
{\small VIDE} is based on the {\small ZOBOV}  algorithm \citep{Neyrinck_etal_2008}, which allows to identify
wells in the density field produced by a set of points. This is done by means of a
a Voronoi tessellation scheme that associates a cell to each tracer of the density field
and subsequently identifies local density minima among these
cells by looking for cells with a larger Voronoi volume than all surrounding cells.
By joining together the Voronoi cells around a
local density minimum based on the Watershed Transform algorithm \citep[see][]{Platen_etal_2007},
a hierarchy in the structures of the identified voids is naturally obtained.
Finally, the {\small VIDE} toolkit elaborates on the void catalog obtained by {\small ZOBOV}
by performing various possible selections of the void sample as e.g. different cuts on the void density contrast or on the void central overdensity.

For our analysis we will consider only voids with a central density below $20\%$ of the mean density of the universe and with a density contrast between the most underdense particle of the void and the void boundary no lower than $1.57$, corresponding to a probability that the void arises from Poisson noise below $\sim5\%$ \citep[see][]{Neyrinck_etal_2008}. 

A standard quantity to characterise cosmic voids and study their relative abundance
is given by their effective radius $R_{\rm eff}$, defined
from the Voronoi volume of the void as the radius of a sphere having the same volume as the void:
\begin{equation}
V_{\rm VOID} \equiv {\sum\limits_{i=1}^N V^{p}_i }= \frac{4}{3} \pi R_{\rm eff}^3 \,.
\end{equation}
However, following the discussion presented in \citet{Yang_etal_2015}, for the analysis presented in this work it is more convenient to characterise voids based on their core density, defined as the  density of the most underdense Voronoi cell of each void. Such quantity will be employed in Section~\ref{sec:voids_abundance} below to characterise the abundance of voids in the different cosmologies under investigation.

\section{Results}
\label{sec:results}

We now present the main outcomes of our simulations. The various observables that will be discussed in this section
have been extracted from the simulations snapshots using standard and well-tested numerical pipelines that are routinely
employed for the analysis of cosmological simulations, and are presented in a very standard form. Therefore, we will not describe very extensively these procedures, the novelty of the present work being in the cosmological models investigated rather than on the analysis performed. We will focus on the statistical and structural properties of both dark matter halos and cosmic voids.

\subsection{Large-scale matter distribution}

\subsubsection{The matter power spectrum}

\begin{figure}
\includegraphics[width=\columnwidth]{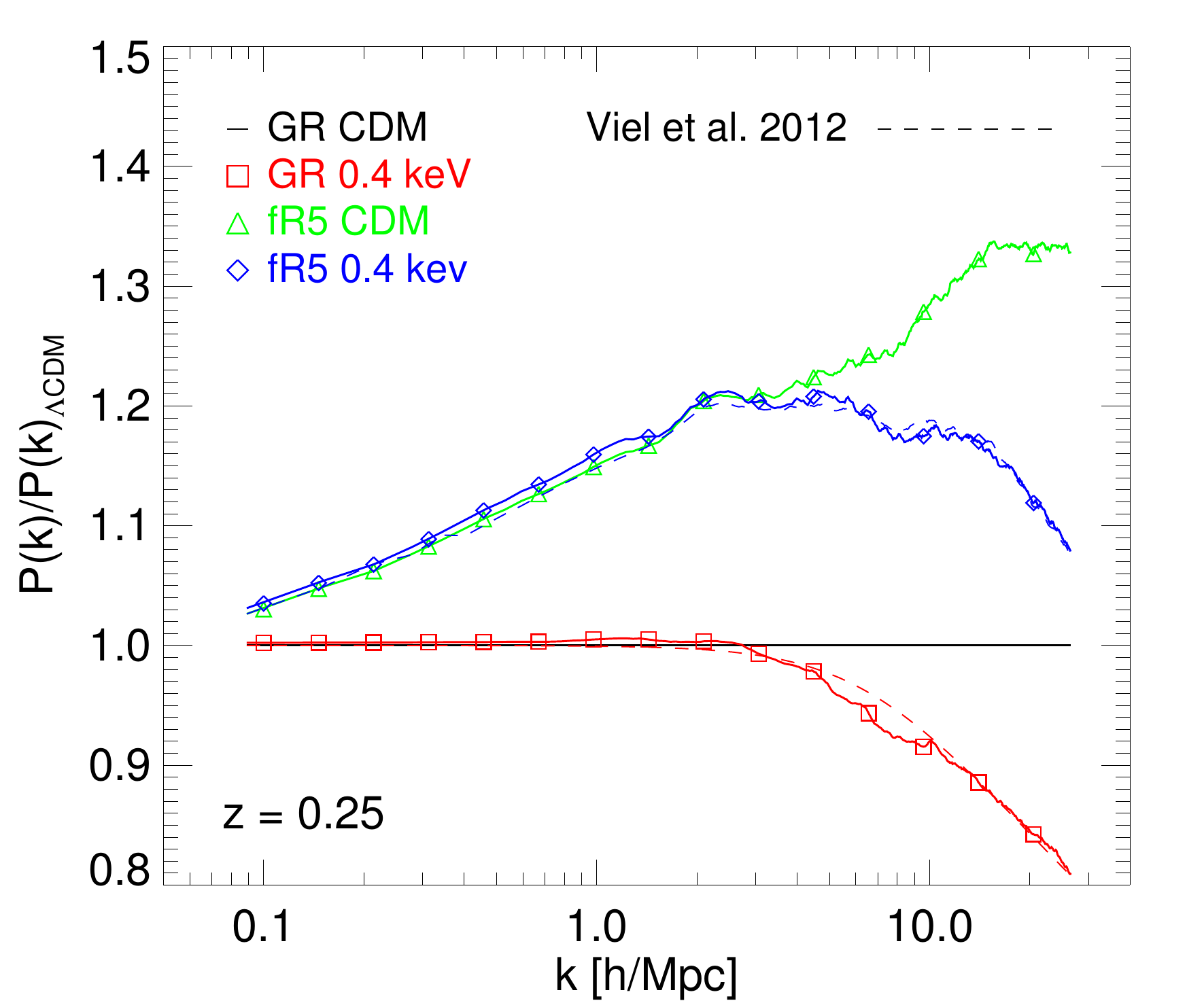}
\caption{The ratio of the non-linear matter power spectrum to the reference model for the different cosmologies under investigation.}
\label{fig:Pk}
\end{figure}

For all the simulations presented above we have computed the matter power spectrum at $z=0.25$ through a Cloud-in-Cell mass assignment to a cartesian grid with $768^{3}$ nodes,
thereby spanning the range of Fourier modes between $k_{0}=0.06\, h/$Mpc and the Nyquist frequency of $k_{\rm Ny}\approx 24\, h/$Mpc. In Figure~\ref{fig:Pk} we show the ratio of the nonlinear matter power spectrum at $z=0.25$
to the standard cosmological scenario in the various models under investigation. As one can see from the plot, the WDM simulation within standard GR (solid red curve with open squares) shows the expected suppression of power at small scales, with deviations from the standard $\Lambda $CDM case starting at the cut-off scale $k_{\rm c}\approx 2\, h/$Mpc. As a reference, we overplot as a dashed line the result of the nonlinear fitting function provided by \citet{Viel_etal_2012}:
\begin{eqnarray} \label{eq_fitting}
&&T^2_{\rm nl}(k)\equiv P_{\rm WDM}(k)/P_{\rm \Lambda CDM}(k)=(1+(\alpha\,k)^{\nu l})^{-s/\nu} \nonumber , \\
&& \alpha(m_{\rm WDM},z)=0.0476\,\left(\frac{1 \rm{keV}}{m_{\rm WDM}}\right)^{1.85}\,\left(\frac{1+z}{2}\right)^{1.3}, \nonumber  \\
&& 
\end{eqnarray}
where we have used the values $\nu=3$, $l=0.8$ and $s=0.2$ for the fitting parameters. As the figure shows, this analytical formula provides an excellent fit to the power suppression in the standard GR case.

The $f(R)$ simulation for the case of CDM particles (solid green curve with triangles) gives rise to the expected scale-dependent power enhancement due to the action of the fifth-force associated with the extra scalar degree of freedom $f_{R}$. This is consistent with a number of previous works and in particular with the outcomes of the code comparison project for $f(R)$ gravity cosmologies presented in \citet{Winther_etal_2015}.

When we consider the combined simulation for a WDM particle with mass $m_{\rm WDM}=0.4$ keV and $f(R)$ gravity (solid blue curve with diamonds) we find (quite expectedly) that at large scales no significant difference appears with respect to the CDM-$f(R)$ case, while at scales below the cut-off $k_{c}$ the power is suppressed compared to the CDM case. This is qualitatively expected, but the present work provides the first quantitative determination of the power suppression due to WDM in the context of a modified gravity cosmological model. In particular, it is remarkable to notice that the nonlinear suppression relative to the CDM model within $f(R)$ gravity shows the same shape and amplitude that is found for the GR case: the blue dashed curve in Figure~\ref{fig:Pk} -- which represents the suppression obtained by applying the best-fit suppression function from Eq.~(\ref{eq_fitting}) for the GR case to the $f(R)$-CDM simulation -- very well captures the power spectrum ratio computed from the simulations. 

\begin{figure}
\includegraphics[width=\columnwidth]{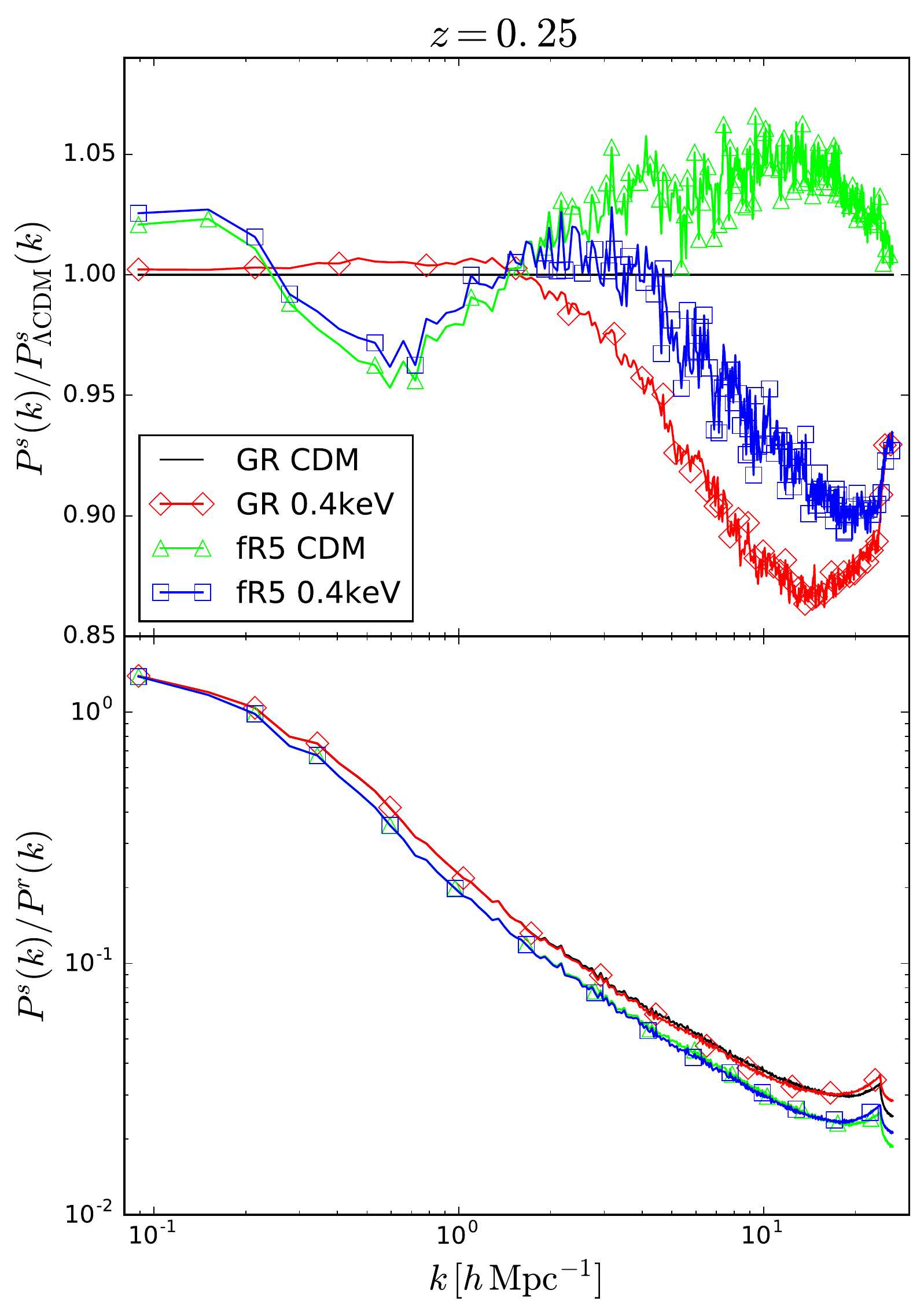}
\caption{{\em Upper panel}: Monopole of the matter power spectrum in redshift-space normalized by the monopole of the $\Lambda $CDM model. {\em Lower panel}: Ratio between the monopoles in redshift- and real-space for the four different models. All results are shown at $z=0.25$.}
\label{fig:RSD}
\end{figure}
\ \\

We have also computed the monopole of the matter power spectrum in redshift-space at $z=0.25$ for the four different models we investigate in this work. We move the particles from real-space to redshift-space using the distant observer approximation. We obtain three different realizations in redshift-space by using the peculiar velocities of the particles along the three different cartesian axes. The monopole in redshift-space is obtained by averaging over the results from each realization. In Fig.~\ref{fig:RSD} we show the results. 

The upper panel of Fig.~\ref{fig:RSD} displays the matter monopole in redshift-space, normalized by the monopole of the $\Lambda$CDM model. Differences among all models are much smaller in redshift-space than in real-space. The model with modified gravity and CDM is degenerate with $\Lambda$CDM at the $\sim 5\%$ level. The models with WDM exhibit a suppression of power on small scales, with respect to their CDM counterparts. We find that the wavenumber were the cut-off shows up in the WDM models depends on whether the model is with GR or modified gravity. 

In the bottom panel of Fig.~\ref{fig:RSD} we plot the ratio between the monopoles in redshift- and real-space for the four different models. As can be seen, there is a clear distinction between models with GR and modified gravity, since the shape and amplitude of the ratio are different on small scales. On the other hand, WDM does not seem to produce any effect on the monopoles ratio. On large-scales, the monopoles ratio is expected to approach the value:
\begin{equation}
\lim_{k \to 0} \frac{P^s(k)}{P^r(k)} = 1+\frac{2}{3}\beta+\frac{1}{5}\beta^2~,
\end{equation}
where the redshift-space distortion parameter is $\beta=f(z)/b$, with $f(z)$ being the linear growth factor and $b$ is the bias (in this case $b=1$). Unfortunately, our simulation boxes are too small to reach the above Kaiser limit. 

\subsubsection{The halo bias}

\begin{figure}
\includegraphics[width=\columnwidth]{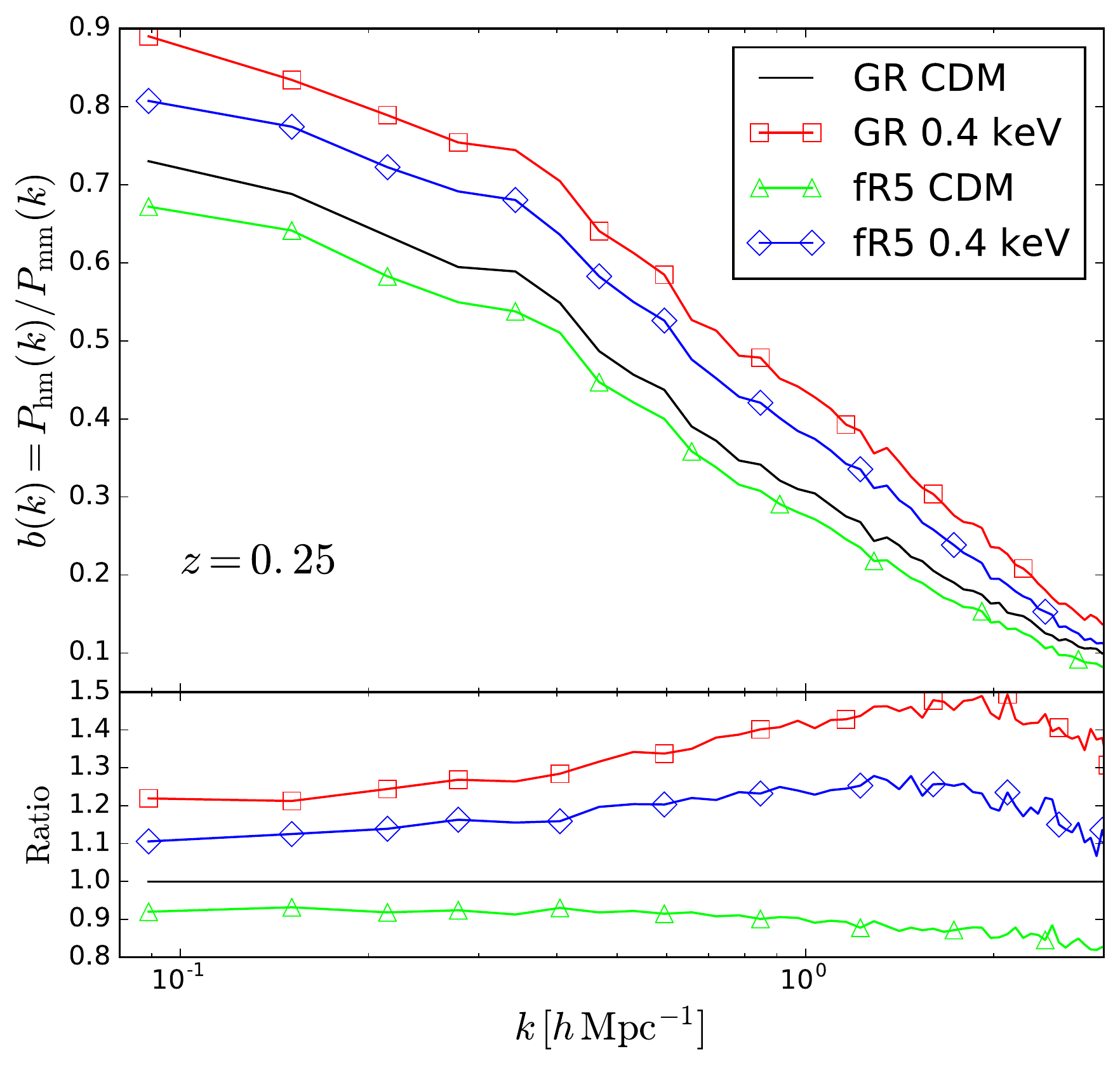}
\caption{Halo bias at $z=0.25$ in real-space. Upper panels shows the halo bias, computed as the ratio of the halo-matter cross-power spectrum over the matter auto-power spectrum for the different models, as indicated in the legend. The bottom panel displays the halo bias normalized by the results of the $\Lambda $CDM cosmology.}
\label{fig:bias}
\end{figure}

We have also investigated the clustering properties of dark matter halos in the different models using the halo bias as our cosmological statistics. We have computed for each model the halo bias as the ratio between the halo-matter cross-power spectrum to the matter auto-power spectrum. We have used this definition of halo bias to avoid stochasticity. Results are shown in Fig. \ref{fig:bias}.

Unfortunately, our simulation boxes are not large enough to allow us to explore the linear bias and whether any of these models exhibit a scale-dependent bias on large scales. We find that models with WDM present a higher amplitude of the halo bias than their CDM counterparts, irrespective of the underlying gravity model. This is the expected effect of WDM \citep{Maccio_etal_2012, Dunstan_etal_2011}: suppressing the abundance of low mass halos and therefore enhancing the clustering of halos of all masses. 

The halo bias of the modified gravity models is systematically lower than the one of the standard GR models by $\sim10\%$. The bottom panel of Fig. \ref{fig:bias} shows the halo bias of the different models normalized by the results of the $\Lambda $CDM model. We find that on the largest scales the relative differences among the various models are almost scale-independent.

\subsection{Statistical and structural properties of Dark Matter halos}
\label{sec:halos}

\subsubsection{Halo Mass Function}

\begin{figure}
\includegraphics[width=\columnwidth]{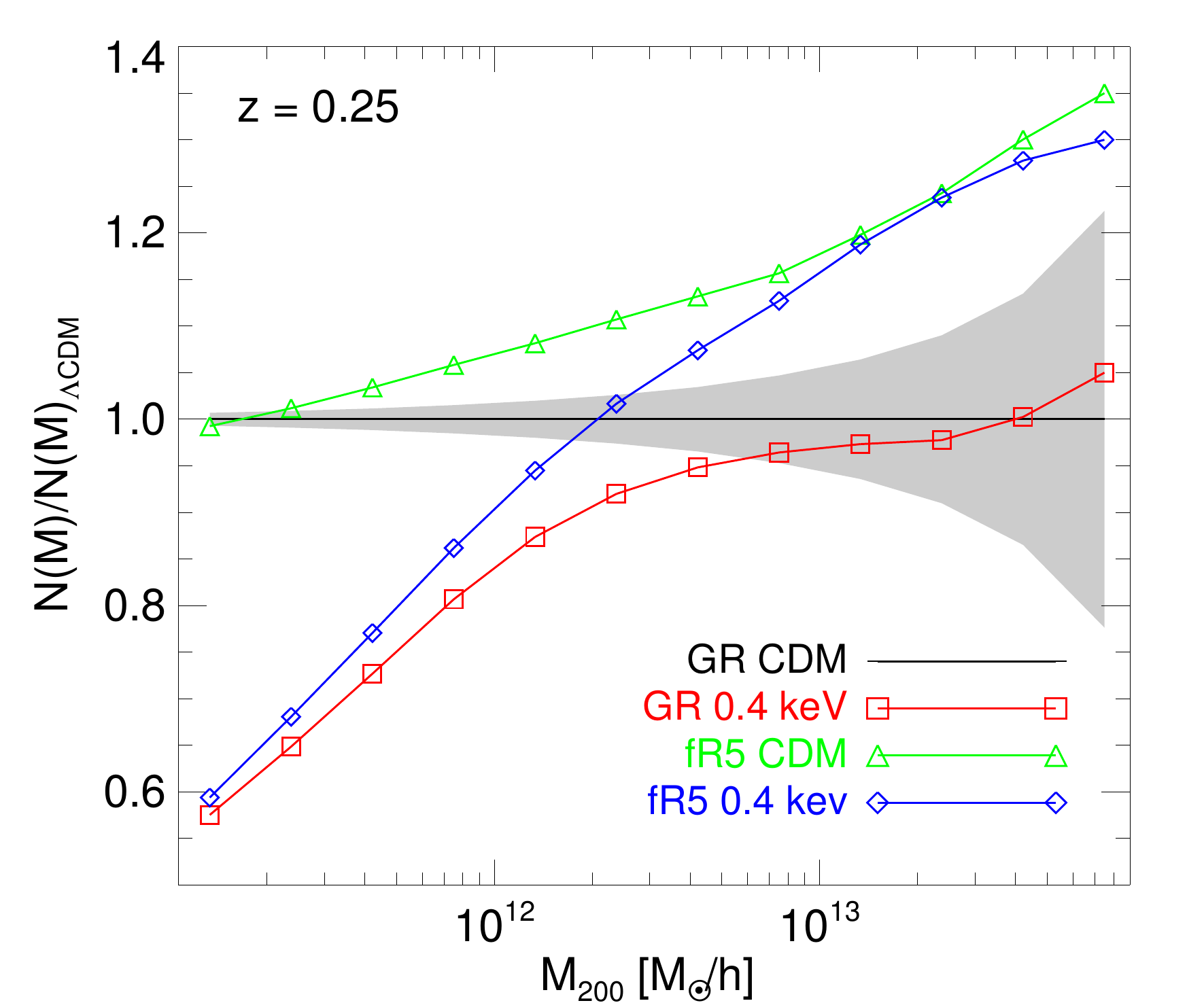}
\caption{The halo mass function for the three cosmologies under investigation. The grey-shaded region represent the Poissonian error propagated to the counts ratio based on the number of halos in each bin.}
\label{fig:HMF}
\end{figure}

We compute the differential halo mass function of all our cosmologies as the number of halos identified by the {\small SUBFIND} algorithm having a virial mass $M_{200}$ (see above) falling within 12 logarithmically equispaced mass bins ranging between $10^{11}$ and $10^{14}\, M_{\odot }/h$. In Fig. ~\ref{fig:HMF} we display the ratio of the abundance computed in each bin to the fiducial case represented by the $\Lambda $CDM cosmology. 

As expected, the WDM model (shown as red open squares in the plot) results in a strong suppression of the abundance of small halos with a reduction of $\approx 40\%$ of objects at the smallest mass side of our considered interval. At larger masses the halo abundance of the WDM cosmology recovers the $\Lambda $CDM expectation within statistical errors, displayed as the grey-shaded area in Fig.~\ref{fig:HMF} and corresponding to the Poissonian error in each bin.

The $f(R)$ cosmology with CDM particles (blue open diamonds) shows on the contrary a mass-dependent enhancement of the abundance of halos that becomes more pronounced for larger halo masses, reaching a $\approx 35\%$ increased halo abundance at the largest mass available to our halo sample. This is also an expected result and fully consistent with the outcomes of several previous works \citep[as e.g., ][just to mention some]{Ecosmog,Puchwein_Baldi_Springel_2013, Lombriser_etal_2013, Winther_etal_2015, Achitouv_etal_2016}. 

By comparing these first two simulations it  appears already clear that no degeneracy  between WDM and $f(R)$ gravity is to be expected for an observable as the halo mass function, since the two scenarios give rise to the opposite mass-dependence of the deviation from the standard cosmological scenario. This is confirmed by directly checking the combined simulation (green open triangles) which shows how the small-mass suppression and the large-mass enhancement both persist in the combined halo mass function. Therefore, the combination of WDM and $f(R)$ gravity is expected to enhance the signal of a mismatch between the detected abundance of halos at large and small masses.

\subsubsection{Subhalo Mass Function}
\begin{figure}
\includegraphics[width=\columnwidth]{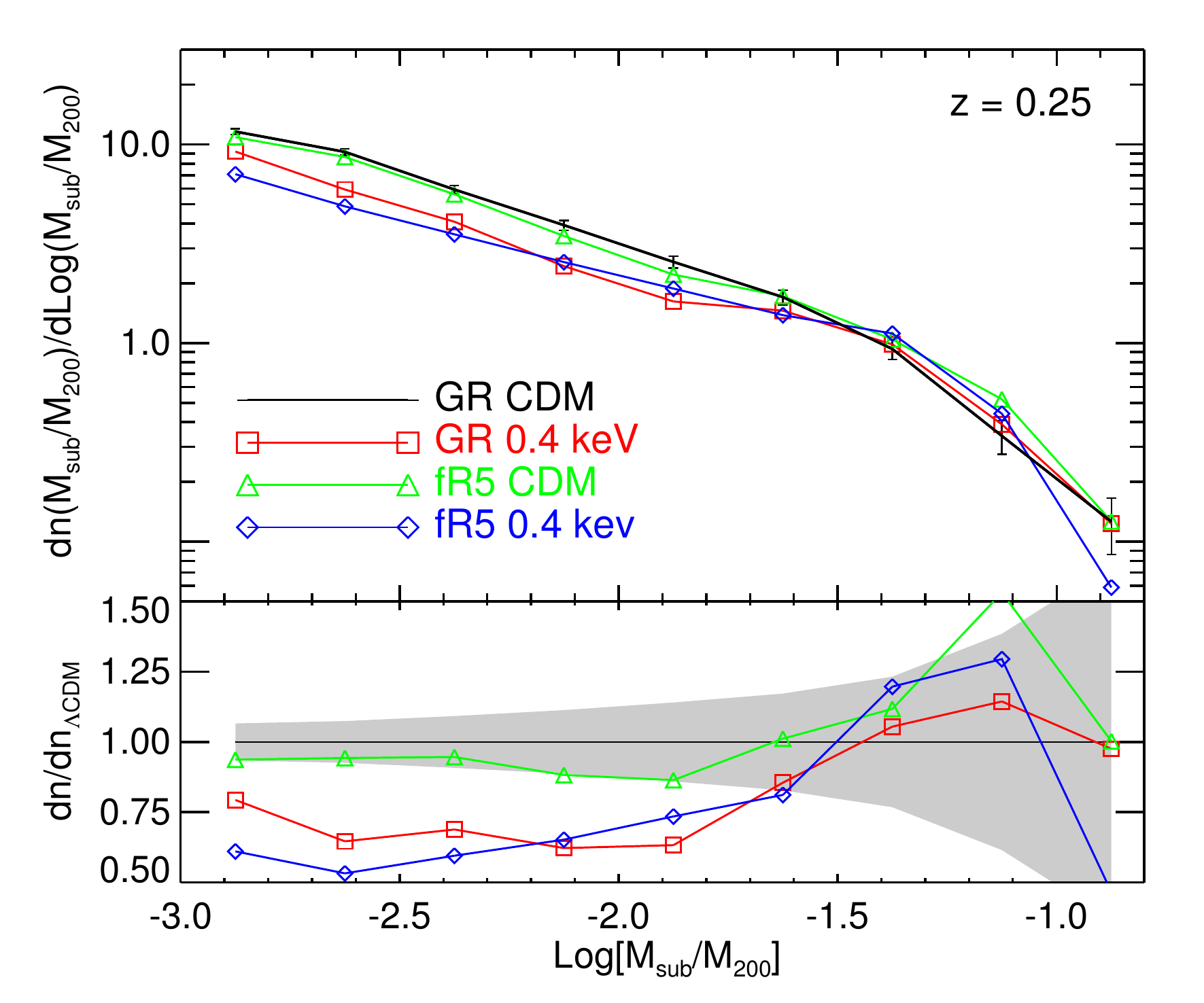}
\caption{The subhalo mass function for the three cosmologies under investigation. The grey-shaded region represent the Poissonian error propagated to the counts ratio based on the number of subhalos in each bin.}
\label{fig:subHMF}
\end{figure}

We also compute -- for all the cosmologies under investigation -- the subhalo mass function, defined as the abundance of substructures of mass $M_{\rm sub}$ that are gravitationally bound to a main halo of virial mass $M_{200}$, as a function of the mass ratio $M_{\rm sub}/M_{200}$. The results are displayed in Fig.~\ref{fig:subHMF} with the same color- and symbol-coding of the previous figures.

As one can see from the plot, the WDM case shows the expected suppression of the abundance of substructures, with about $25-40\%$ fewer subhalos compared to $\Lambda $CDM for a mass fraction $M_{\rm sub}/M_{200}$ below $10^{-2}$. The effect is significant when compared to the statistical error due to Poisson noise, which is represented in the plot by the grey-shaded region.

This suppression does not appear to be significantly modified when moving from GR to $f(R)$ as the underlying theory of gravity. In fact, the combined simulation shows approximately the same behavior as the GR-WDM run. On the contrary, the $f(R)$-CDM simulation -- for which no suppression of the abundance of substructures is expected -- is found to be consistent with the standard scenario within statistical errors.

Therefore, also for the case of the abundance of bound substructures within virialised halos no significant degeneracy is found between the effects of a WDM particle and a modified theory of gravity in the form of $f(R)$.

\subsubsection{Halo density profiles}
\label{sec:halo_profiles}

\begin{figure*}
\includegraphics[width=\textwidth]{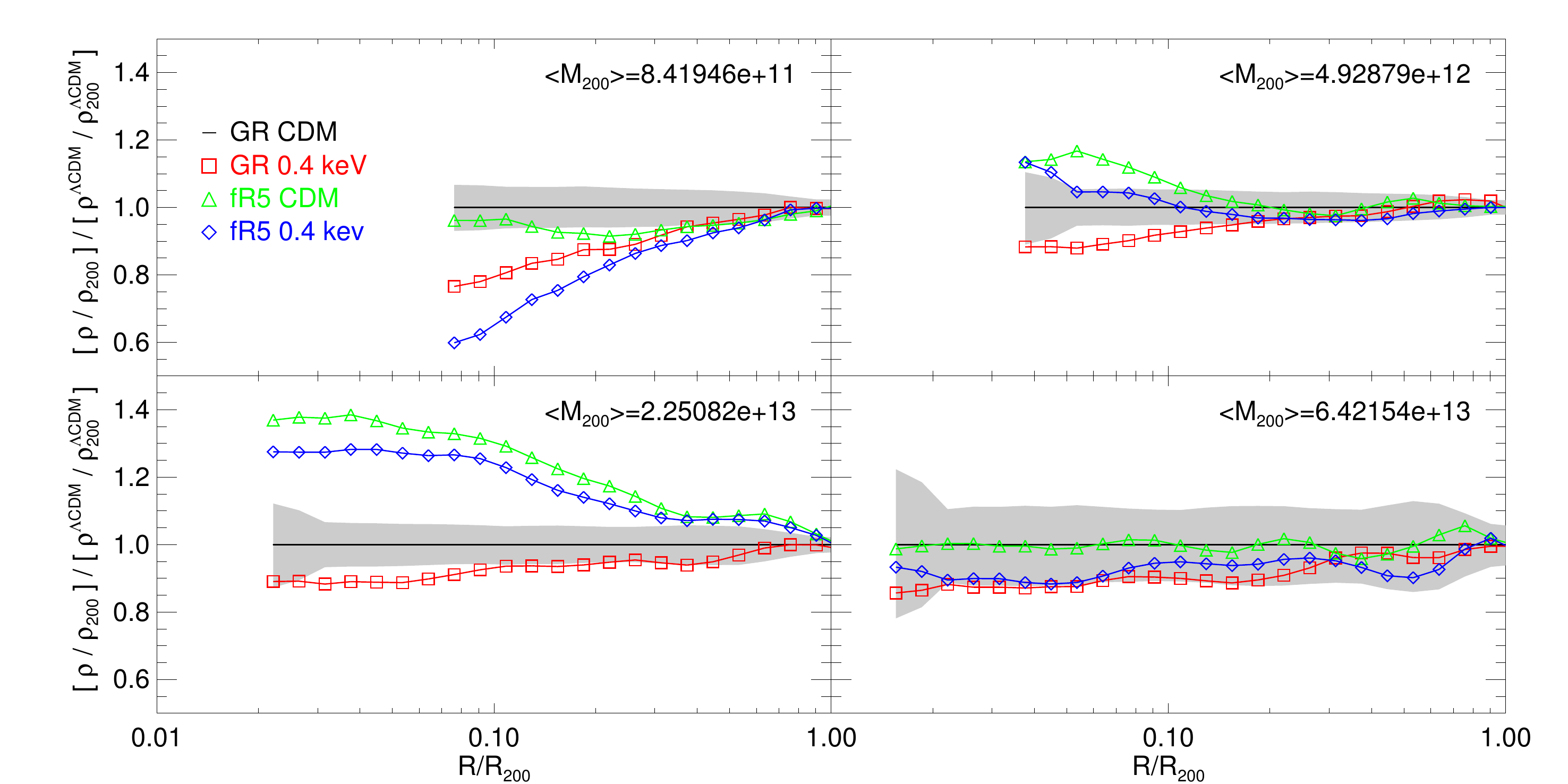}
\caption{The ratio of the stacked halo density profiles to the fiducial GR+CDM cosmology for about 100 halos within 4 out of the 12 mass bins considered for the halo mass function and concentration analysis. The grey shaded area represents a $2-\sigma $ uncertainty on the stacked profile based on a bootstrap procedure.}
\label{fig:halo_profiles}
\end{figure*}

For each of the 12 mass bins employed to compute the differential halo mass function discussed above we have computed the average density profiles of up to 100 halos. This means that we have computed the mass density in a set of 30 logarithmic radial shells centered on the most bound particle of a halo for a random sample of halos belonging to each mass bin, and then stacked these density profiles after rescaling the individual radial coordinates in units of the halo virial radius $R_{200}$ to obtain an average profile for each bin. For those bins not reaching 100 members (normally the last three) we have performed the stacking using all the available members. 

All the profiles, independently on the mass bin and on the cosmological model, have then been normalised to unity at the virial radius and we show the ratio of these profiles to the $\Lambda $CDM cosmology in Fig.~\ref{fig:halo_profiles} for four out of the twelve available mass bins. We restrict the analysis to masses above $\approx 8\times 10^{11}\, M_{\odot }/h$ as the lower mass halos are poorly resolved and the resulting profiles are noisy. The grey-shaded areas in the plot represent the 2-$\sigma $ confidence limits based on the standard deviation of the mean density profiles computed through a {\em bootstrap} resampling technique with 1000 re-samples of the 100 individual profiles. 

As one can see in the plots, the WDM realisation for a standard GR theory of gravity (red open squares) always results in  shallower density profiles with a significant suppression of the central overdensity and a lower profile slope, even though the effect is -- as expected -- more pronounced for the lowest mass bin and progressively decreases for higher-mass halos. In particular, for the largest masses available to our sample the profile is marginally consistent with the standard cosmological result at $\approx 2\sigma $. This is the very well known effect of the thermal cut-off in the primordial density power spectrum, which determines the formation of cored density profiles for low mass halos and of progressively steeper profiles for increasing halo mass \citep[][]{Villaescusa-Navarro_Dalal_2011,Maccio_etal_2012}.

On the contrary, the CDM realisation for $f(R)$ gravity (green open triangles) shows a more complicated modulation of the deviation with respect to the standard model as a function of the halo mass. More specifically, while the profiles are only mildly affected at the lowest mass bin and completely unaffected at the highest mass bin displayed in the plots, the effect is more pronounced for intermediate masses and results in a significant steepening of the profiles with up to a $40\%$ enhancement of the central overdensity for halos of mass $\approx 1-2 \times 10^{13}\, M_{\odot }/h$. This behaviour is consistent with the effects of the {\em Chameleon} screening mechanism that characterises $f(R)$ gravity, with the most massive halos being fully screened due to their deep potential wells, and the lowest mass halos having formed and virialised at earlier epochs when the higher average cosmic density significantly suppresses the scalar fifth force.

It is particularly interesting then to test the outcomes of the combined WDM-$f(R)$ simulation (open blue diamonds) where the interplay between these opposite effects and their different dependence on the host halo mass might give rise to characteristic observational footprints. In particular, we find (quite surprisingly) that the shallowing of the density profiles for the low-mass halos presented in Figure~\ref{fig:halo_profiles} is significantly enhanced in $f(R)$ gravity compared to standard GR. On the contrary, for intermediate masses (top-right and bottom-left plots in the Figure) the effect of the fifth-force seems to overcome the suppression due to WDM and the resulting density profiles are steeper than in the standard cosmological scenario, even though with a weaker enhancement of the central overdensity as compared to the CDM case.
Finally, for the most massive halos considered in our analysis the effect of the fifth-force seems to be completely absent (due to the {\em Chameleon} screening, as discussed above) and the combined model appears fully consistent with the WDM result in standard GR.

\subsubsection{Concentrations}

\begin{figure*}
\includegraphics[width=0.45\textwidth]{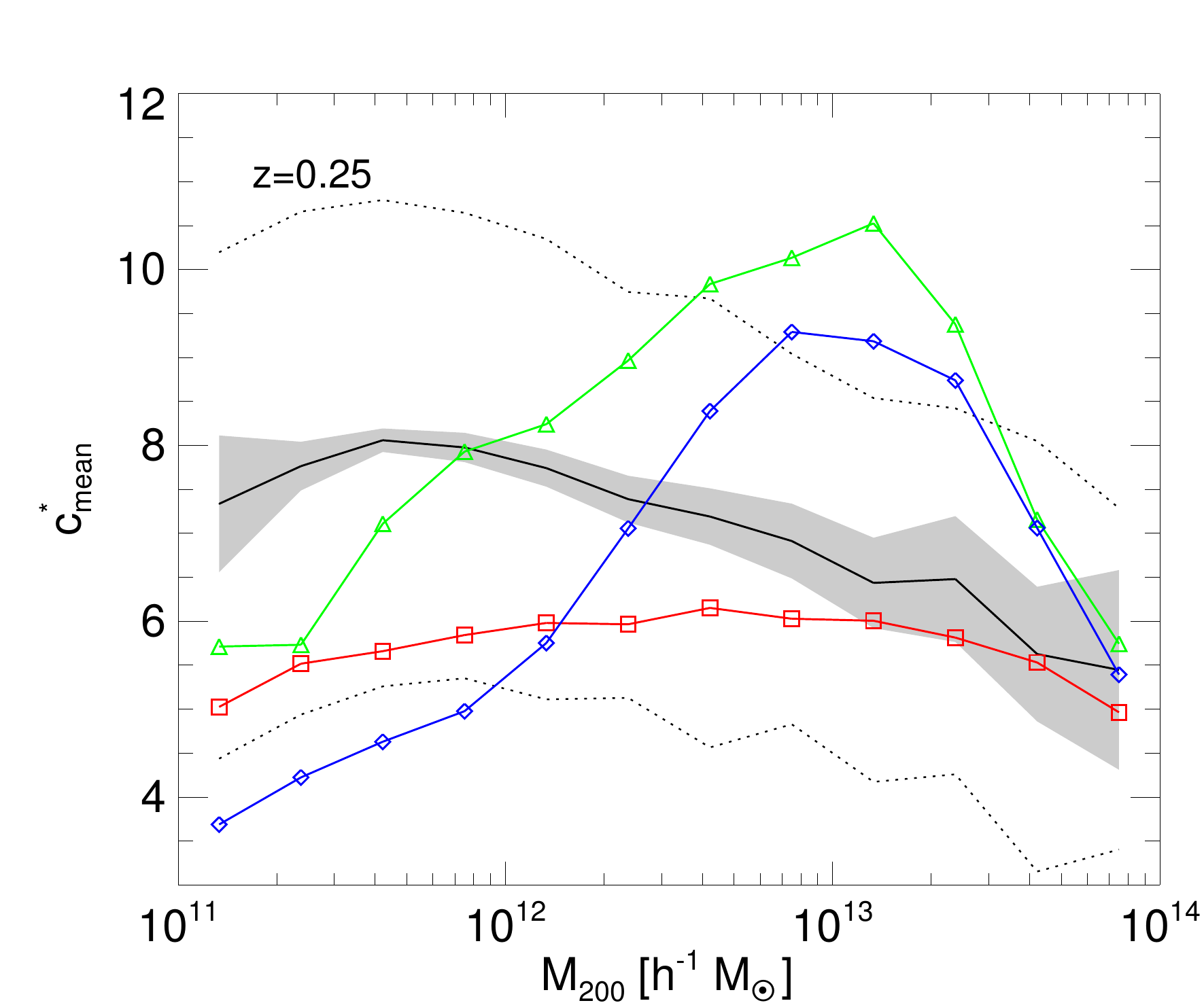}
\includegraphics[width=0.45\textwidth]{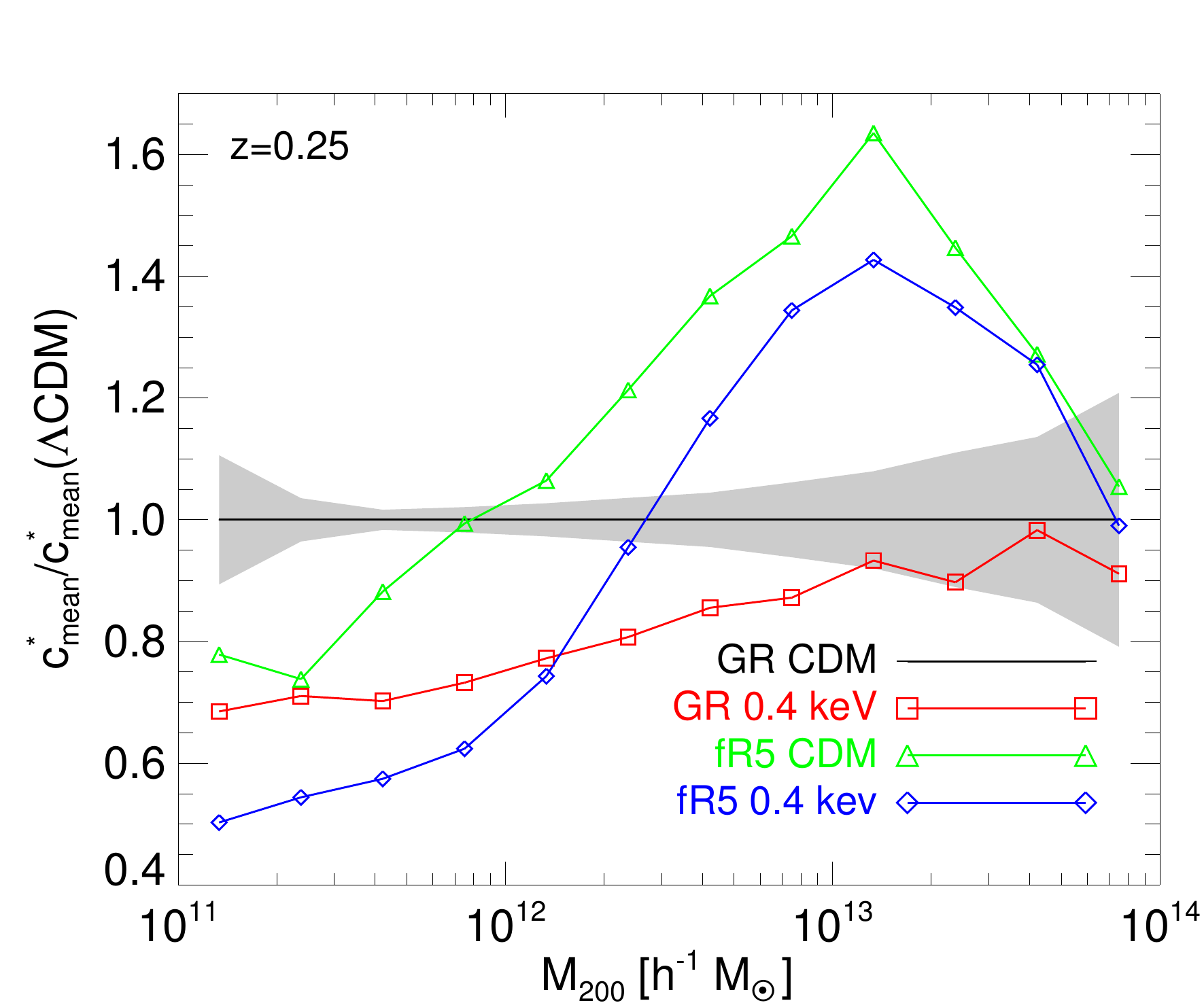}
\caption{The concentration-mass relation ({\em Left}) and its ratio to the standard cosmological model ({\em Right}) for the three cosmologies under investigation. The grey-shaded regions represent the Possonian errors on both the concentrations ratio and the mean concentration values due to the number of halos used in each mass bin. The dotted lines in the {\em Left} plot indicate the spread of 68\% of the halos in each bin.}
\label{fig:concentrations}
\end{figure*}

For each halo in our sample we have computed the concentration $c^{*}$ following the approach described in \citet{Aquarius} as:
\begin{equation}
\label{conc_vmax}
\frac{200}{3}\frac{c^{*3}}{\ln (1+c^{*}) - c^{*}/(1+c^{*})} = 7.213~\delta _{V}
\end{equation}
where $\delta _{V}$ is defined as:
\begin{equation}
\delta _{V} = 2\left( \frac{V_{max}}{H_{0}r_{max}}\right) ^{2}
\end{equation}
with $V_{max}$ and $r_{max}$ being the maximum rotational velocity of the halo and the radius at which this velocity peak is located, respectively. Both quantities have been computed in all models based only on the gravitational potential associated with the matter density profile, therefore neglecting the effect of the additional fifth-force. This approach to compute halo concentrations has already been employed in the context of $f(R)$ modified gravity theories by \citet{Arnold_Puchwein_Springel_2016} for a small set of very well resolved Milky-Way-sized halos (by means of a set of zoomed high-resolution simulations) and compared with the results obtained from fitting the individual density profiles with a Navarro-Frenk-White \citep[NFW][]{NFW} shape. In our setup we have a much poorer resolution than in the simulations of \citeauthor{Arnold_Puchwein_Springel_2016} but a significantly larger statistics. We have therefore repeated the comparison by performing a $1$-parameter fitting of an NFW shape for the individual  profiles of all the randomly-selected halos used to compute the stacked density profiles discussed in the previous section, and compared these determinations of the average concentration in each of the 12 mass bins adopted for both the halo mass function and the halo density profiles with the one obtained from Eq.~(\ref{conc_vmax}). We found  consistent results on the relative effects of the different models between these two different ways to compute concentrations for the 5 most massive bins (for which the density profiles are reasonably well resolved and the fitting procedure is not exceedingly affected by the radial fitting range). For lower mass halos the agreement is significantly worse, but this is to be expected due to the poor resolution of the individual profiles and the large scatter in the fitted parameters. In the following we will then assume that the concentration values provided through Eq.~(\ref{conc_vmax}) are reliable over the whole mass range covered by our sample and we will show only these results in the following. This assumption would clearly require a more detailed scrutiny though a sample of simulated halos with comparable statistics but higher resolution than allowed by our simulations. 

In Figure~\ref{fig:concentrations} we show the concentration-mass relation ({\em left} panel) and its ratio to the standard cosmological model ({\em right} panel) for all the cosmologies under investigation.
The grey-shaded areas represent the 2-$\sigma $ confidence regions based on the Poisson errors on the mean for each mass bin, while the dotted curves in the {\em left} panel show the spread of 68\% of halos around the mean in each bin.

As one can see from the two plots, the WDM model in standard GR shows the expected behaviour of suppressing halo concentrations at low masses and progressively recovering the standard $\Lambda $CDM concentrations at larger masses. A maximum suppression of $\approx 30\%$ is found for the lowest masses available to our sample, while at the four largest mass bins the model is again consistent with $\Lambda $CDM at 2-$\sigma $.

The CDM model within $f(R)$ gravity, instead, shows a significant increase of halo concentrations for intermediate masses and recovers the standard value of the concentration-mass relation only at the highest mass bin. Interestingly, the lowest mass halos with virial masses $M_{200} \lesssim 2\times 10^{12}$ M$_{\odot }/h$ show a lower concentration compared to the standard model. This is qualitatively consistent with the behaviour of the density profiles shown in Figure~\ref{fig:halo_profiles} above. Such suppression of halo concentrations has been found for a fraction of the considered halos with masses around $2\times 10^{12}$ M$_{\odot }/h$ also in the higher-resolution study of \citet{Arnold_Puchwein_Springel_2016} and for a lower value of the $f(R)$ scalar amplitude ($\bar{f}_{R0}=-10^{-6}$), while the majority of halos did show an increased concentration. Again, a resolution comparable to \citeauthor{Arnold_Puchwein_Springel_2016} for a halo sample with similar statistics to our work would be required to robustly assess the impact of $f(R)$ gravity on the concentrations of halos below such mass value.

Finally, we investigated the outcomes of the combined WDM-$f(R)$ cosmology which shows a very interesting distortion of the concentration-mass relation, again fully consistent with the results obtained for the density profiles: a weaker enhancement of halo concentrations compared to the CDM realisation of $f(R)$ gravity for intermediate masses with a {\em steeper} and {\em stronger} suppression of halo concentrations at low masses even when compared with the WDM cosmology for standard GR. 

The suppression of the inner halo density -- and consequently of the halo concentration -- for low mass halos that we find for the $f(R)$ cosmologies for both CDM and WDM compared to the standard GR case (Figures~\ref{fig:halo_profiles} and \ref{fig:concentrations}) is not easily understandable in terms of the different regimes of action of the {\em Chameleon} screening mechanism and would deserve further investigation, in particular -- as already stated above -- a more detailed analysis based on higher-resolution simulations, which we defer to future work.

\subsection{Statistical and structural properties of cosmic voids}

Cosmic voids have attracted significant interest in the past few years as possible complementary cosmological probes to the standard large-scale structure observables. In particular, previous investigation of the statistical and structural properties of cosmic voids in modified gravity theories have been performed by e.g. \citet{Cai_Padilla_Li_2015}, while a study of the impact of WDM on cosmic voids has been presented in \citet{Yang_etal_2015}. 

In the present work we investigate for the first time the joint effects of these two modifications of the standard cosmological model on cosmic voids, focusing in particular on the abundance and on the density profiles of relatively small voids. Although a realistic determination of the expected observational properties of cosmic voids should properly take into account the effect of the bias of the visible tracers (such as galaxies and clusters) employed to identify the voids in large cosmological surveys \citep[see e.g.][]{Pollina_etal_2016,Nadathur_Hotchkiss_2015}, a clear understanding of the properties of cosmic voids in the underlying matter density field represents an essential step to employ voids as cosmological probes.

\subsubsection{Abundance of cosmic voids}
\label{sec:voids_abundance}

\begin{figure}
\includegraphics[width=\columnwidth]{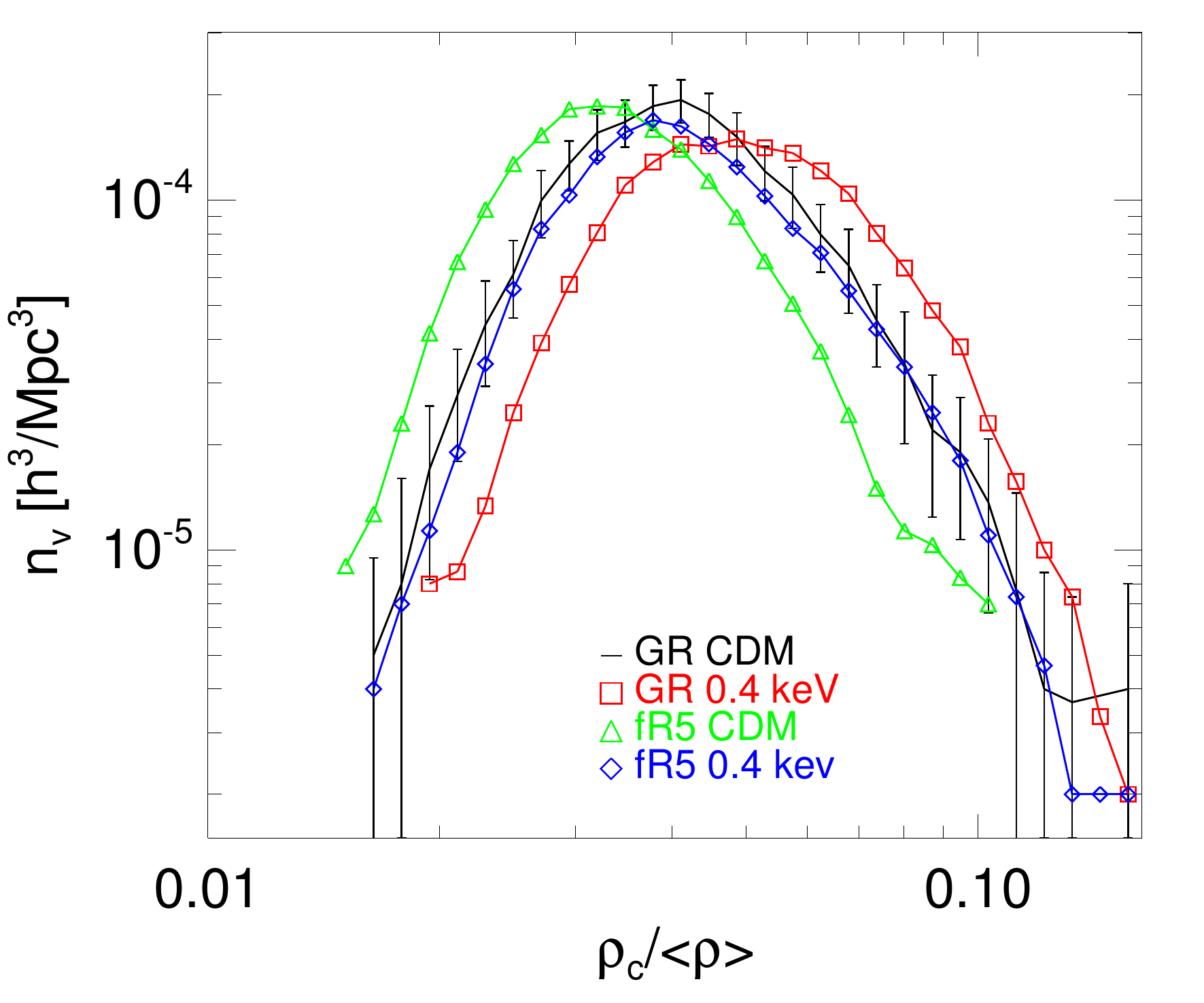}
\caption{The differential distribution function of cosmic voids  as a function of their core density (i.e. the number density $n_{v}$ of voids in each core density bin) for the various cosmologies. The error bars on the reference $\Lambda $CDM model correspond to the 2-$\sigma $ Poissonian error on the number counts in each bin.}
\label{fig:coredens_dist_func}
\end{figure}

Following the discussion presented in \citet{Yang_etal_2015} we have computed the abundance of cosmic voids as a function of their core density within 30 bins spanning the range of core densities covered by our void sample -- after performing the selection procedures described above in Section~\ref{sec:voids} -- which goes from $0$ to $\approx 0.2$ in units of the mean particle density. Such distribution is in fact more representative of the average structural properties of the void population than the size distribution function, due to the unrealistic assumption of sphericity that enters the definition of the void effective radius \citep[see again][]{Yang_etal_2015}. The resulting core density distribution functions are displayed in Fig.~\ref{fig:coredens_dist_func}, where the error bars overplotted on the standard $\Lambda $CDM curve are 2-$\sigma $errors computed as the Poissonian uncertainties associated with the number counts in each bin. 

As one can see in the plot, consistently with the outcomes of \citet{Yang_etal_2015}, we do find that our WDM model within standard GR  \citep[red open squares, having a WDM mass of $0.4$ keV corresponding to the most extreme cut-off value considered in][]{Yang_etal_2015} determines an overabundance of voids with higher core densities and a corresponding suppression of the abundance of the most empty voids. In this respect, we confirm the results obtained by \citet{Yang_etal_2015} with simulations of comparable resolution as ours. 

The CDM realisation of $f(R)$ gravity (open green triangles), shows exactly the opposite behaviour, with a significant enhancement of the most empty voids and a suppression of the large core-density tail of the distribution.

Quite remarkably, finally, the core density distribution function that results from the WDM model within the $f(R)$  gravity simulation shows that the effect associated with the thermal cut-off due to the low WDM particle mass is almost perfectly counterbalanced by the modified gravitational evolution, so that the resulting abundance of voids is again consistent with the standard model within statistical uncertainties. This result provides the first evidence of an intrinsic observational degeneracy between WDM and $f(R)$ gravity at the level of the statistical properties of the population of small cosmic voids.

Even though the degeneracy between these two modifications of the standard cosmological model is easily broken by a number of other observables (as we have extensively shown in the previous sections) it is interesting to notice that for objects that are only mildly nonlinear (such as cosmic voids) but still at relatively small scales the two models mask each other almost perfectly. This outcome, as we will show in the next section, is not a peculiar coincidence occurring only for the core density distribution function, but appears to characterise also the structural properties of small voids, while for larger voids (arising from larger modes of the primordial density spectrum) the effect of WDM becomes less pronounced so that the expected imprint of $f(R)$ gravity stems out prominently.

\subsubsection{Void density profiles}

\begin{figure*}
\includegraphics[width=\textwidth]{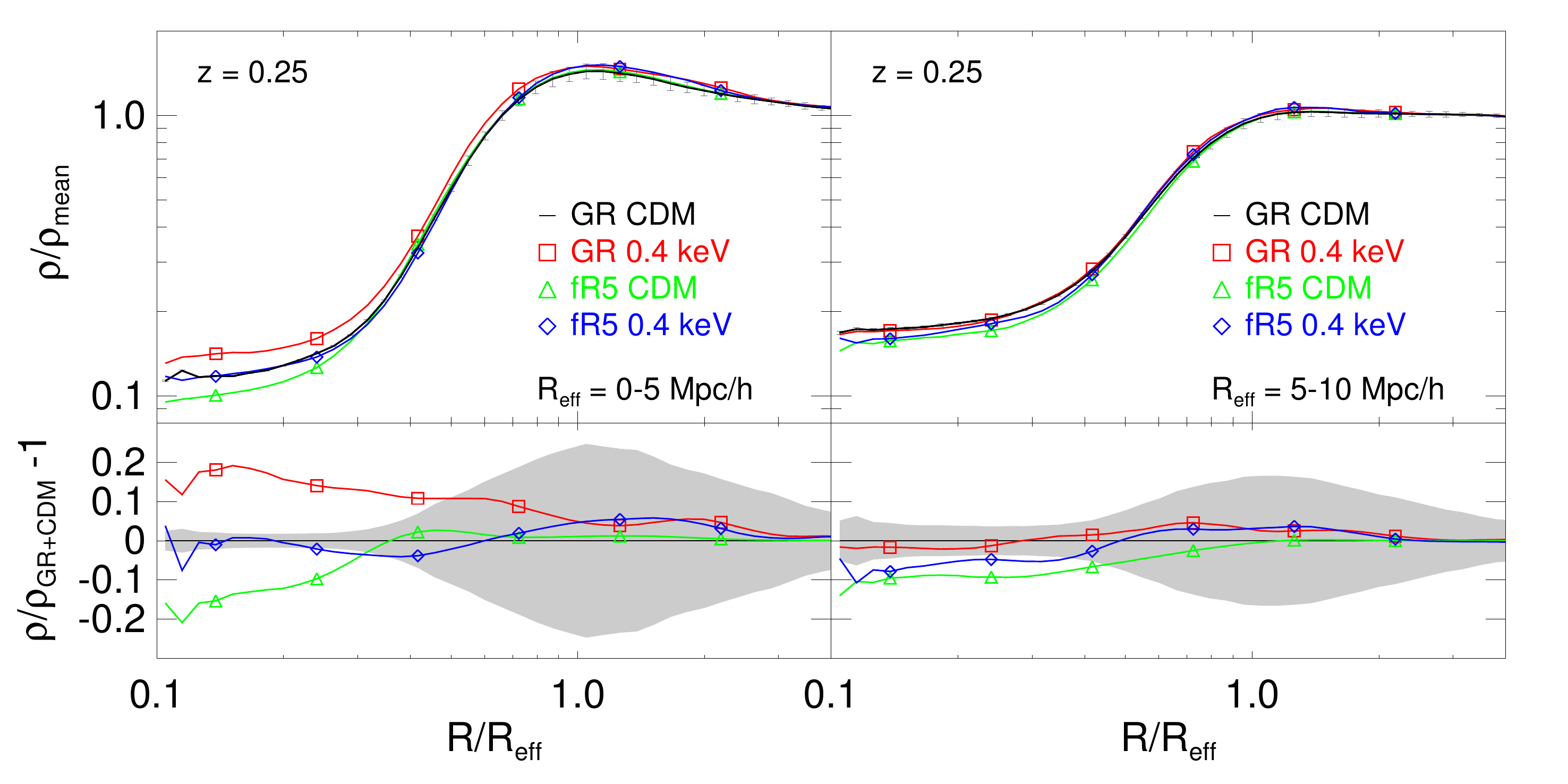}
\caption{The stacked void density profiles in two different ranges of effective radius for the various cosmologies. The {\em upper} plots show the mean spherically-averaged density profiles of 100 randomly selected voids, with the (barely visible) error bars representing the Poissonian error on the mean. The {\em lower} plots display the relative difference with respect to the standard $\Lambda $CDM cosmology with the grey-shaded regions representing the 2-$\sigma $ confidence region based on a bootstrap computation of the standard deviation of the average profiles.}
\label{fig:void_profiles}
\end{figure*}

As a final test of our set of cosmological models we have computed the average void density profiles for voids with effective radius $R_{\rm eff}$ below $5$ Mpc$/h$ and between $5$ and $10$ Mpc$/h$, by stacking the individual spherically-averaged density profiles of 100 randomly selected voids for each of these two bins of $R_{\rm eff}$. The resulting mean profiles are displayed in the {\em left} and {\em right} panels of Fig.~\ref{fig:void_profiles}, respectively. In the two plots, the top panels show the density profiles, with the error bars on the standard $\Lambda $CDM curves representing the Poissonian errors on the mean, while the bottom panels show the relative difference with respect to the standard model, with the grey-shaded regions indicating the 2-$\sigma$ confidence region based on a {\em bootstrap} resampling technique with 1000 re-samples of the 100 individual profiles, as already done for the halo density profiles discussed in Section~\ref{sec:halo_profiles} above. 

Consistently with \citet{Yang_etal_2015} and with the previous results on the core density distribution, we find that WDM within standard GR results in shallower density profiles for the smallest voids, while no statistically significant effect appears for somewhat larger voids.

On the contrary, the effects of $f(R)$ gravity for a standard CDM particle candidate are present for both small and larger voids, resulting in a steeper profile with a lower central density. This is consistent with previous studies on the structural properties of cosmic voids in modified gravity cosmologies \citep[see again][]{Cai_Padilla_Li_2015}.

As a result, the density profiles for the smallest voids in the combined WDM-$f(R)$ cosmology are again fully consistent with the standard $\Lambda $CDM scenario, thereby confirming the degeneracy between these two models in the properties of cosmic voids already identified in the previous section for the core-density distribution. For the larger voids, quite expectedly, the degeneracy is much weaker due to the lower impact of the WDM cut-off and the still significant effect of the scalar fifth-force of the$f(R)$ modified gravity model.

\section{Conclusions}
\label{sec:conclusions}

We have performed an extended analysis of two cosmological N-body simulations featuring either a Warm Dark Matter particle candidate or an $f(R)$ modified gravity theory and compared their outcomes to the standard $\Lambda $CDM cosmology. This analysis has provided fully consistent results with the long series of previous works investigating the same models. 

Additionally, we have presented for the first time the results of a cosmological simulation jointly including the effects of both these deviations from the $\Lambda $CDM cosmology, and tested possible degeneracies of the two independent modifications of the standard model on a wide range of large-scale structure statistics. 

In particular, we have investigated the properties of the large-scale matter distribution by extracting from all our simulations the nonlinear matter power spectrum both in real and redshift space, the statistics of the halo populations within the different models, and the structural properties of collapsed structures over a wide range of masses, as well as the properties of cosmic voids arising  in these different cosmological scenarios. Our results have shown that most observables do not show a significant degeneracy between the effects of the WDM particle and those of a modified law of gravity, while a few other observables are indeed characterised by a strong degeneracy.

More specifically, our main results can be summarised as follows.
\begin{itemize}
\item[$\star $] The {\bf matter power spectrum $P(k)$ in configuration space} shows the well known features for the individual WDM and $f(R)$ cosmologies, i.e. a scale-dependent suppression of power below a cut-off scale $k_{c}\approx 2\, h/{\rm Mpc}$ for the former and an scale-dependent enhancement of power for the latter, both with an amplitude increasing with scale. These effects have been widely discussed in the literature and are well understood as resulting from the primordial cut-off in the density power spectrum and from the effect of the  fifth-force associated with the scalar degree of freedom $f_{R}$, respectively. In particular, for the WDM simulation we find that the cut-off is well captured by the fitting formula provided by \citet{Viel_etal_2012}, which provides a consistency check for our numerical integration. The combined model shows an identical behaviour as the $f(R)$ cosmology down to the cut-off scale $k_{c}$, followed by a suppression of the modified gravity enhancement. Quite remarkably, we find that the suppression induced by WDM with respect to the CDM scenario within $f(R)$ gravity follows the same best-fit shape obtained by the fitting formula of \citet{Viel_etal_2012} for the standard GR simulation. Therefore, our results provides the first validation of this widely-used fitting formula for non-GR cosmologies.
\item[$\star $] The {\bf matter power spectrum $P_{s}(k)$ in redshift space} exhibits differences among models of a maximum of $\sim15\%$, showing that the matter clustering properties are much more degenerate in redshift-space than in real-space. This demonstrates that the effect of peculiar velocities compensates the deviations that the nature of dark matter and the gravity model induce on the clustering of matter. We also find that the ratio of monopoles in redshift and real space clearly segregate models, on small scales, according to the underlying gravity model, while on large-scales the amplitude and shape is very degenerate among all models. 
\item[$\star $] The {\bf halo bias $b(k)$} shows that models with WDM have a larger amplitude, on all scales, with respect to the models with CDM, independently of the underlying gravity theory. We find that on scales $k\lesssim0.2~h{\rm Mpc}^{-1}$ relative differences among models are scale-independent and amount to $\sim20\%$ when comparing cosmologies with the same gravitational theory. Differences in halo bias among models with the same dark matter particle mass but different gravitational theories are somewhat smaller reaching $\sim10\%$.
\item[$\star $] The {\bf halo mass function} shows the well known suppression of the abundance of small halos for the WDM cosmology, and the expected mass-dependent enhancement of the abundance of halos for the $f(R)$ gravity model. In this case, however, differently from the situation encountered for the matter power spectrum, the two effects have the opposite dependence on the halo mass, with the former increasing for progressively lower masses and the latter increasing for progressively larger masses. This different mass dependence ensures that no degeneracy appears when the two models are jointly at work. In fact, the low-mass range of the resulting halo mass function follows the same suppression found for the WDM model while the high-mass range closely follows the $f(R)$ mass function. The transition between these two regimes is found to occur at $\approx 2\times 10^{12}$ M$_{\odot }/h$.
\item[$\star $] The {\bf subhalo mass function} shows a strong suppression of the abundance of substructures, in particular for subhalos with mass ratio to their host main structure below $10^{-2}$, while no statistically significant change in the abundance of substructures is observed for the $f(R)$ gravity scenario. Then, quite expectedly, the combination of the two models substantially follows the behaviour of the WDM cosmology in standard GR with a comparable reduction of the number of small subhalos.
\item[$\star $] The {\bf halo density profiles}, compared among the different models through a stacking procedure of 100 randomly selected halos for several different halo mass bins, shows a significant shallowing of the density profiles for the WDM model at small halo masses, which is progressively reduced for higher mass bins. On the contrary, the $f(R)$ model shows a more complex modulation of the deviation from the standard $\Lambda $CDM profiles, due to the different impact of the {\em Chameleon} screening mechanism for different overdensity environments, with very little deviations observed at the smallest and largest mass bins, and a significant steepening of the profiles for intermediate masses. Interestingly, the combined model is found to enhance the WDM suppression of the inner halo overdensities at small masses and to reduce the $f(R)$ increase of the inner halo overdensities at intermediate masses.
\item[$\star $] The {\bf concentration-mass relation} shows an interesting behaviour, fully consistent with the results obtained for the halo density profiles in the different mass ranges. On one hand, the WDM scenario within standard GR displays the well known suppression of the halo concentrations for low mass halos, while the standard $c-M$ relation is recovered for larger masses; on the other hand, the two $f(R)$ cosmologies for both CDM and WDM are found to provide an enhancement of halo concentrations at intermediate masses, with a peak of the deviation from $\Lambda $CDM at $\approx 10^{13}\, M_{\odot }/h$, and a suppression of concentration at the lowest masses covered by our sample, which is more pronounced for the WDM realisation as compared to the CDM case. While this suppression may be expected in the former case, the deviation found for CDM in $f(R)$ gravity represents an interesting feature deserving further investigation with higher resolution simulations;
\item[$\star $] The {\bf abundance of cosmic voids} is found to be one of the few observables (together with the matter power spectrum in redshift space discussed above and with the void profiles summarised below) for which WDM and $f(R)$ gravity show a significant degeneracy. In particular, as already found by \citet{Yang_etal_2015}, WDM determines a suppression of the abundance of small voids with core density below 5\% of the mean cosmic density, and a corresponding enhancement of the abundance of voids above this threshols. This corresponds to voids being less underdense in WDM compared to CDM. On the other hand, $f(R)$ gravity is found to determine the opposite effect on the population of small cosmic voids, with an enhancement of the abundance of the most underdense voids and a corresponding suppression of the less underdense ones. As a result, the combined cosmology featuring both WDM and $f(R)$ gravity shows a very strong degeneracy with the standard $\Lambda $CDM result, with the core-density distribution function being consistent with the standard model at 2-$\sigma $ confidence level.
\item[$\star $] The {\bf density profiles of cosmic voids} show a different range of effects for the different cosmologies depending on the void size. For small voids ($R_{\rm eff} \leq 5$ Mpc$/h$) WDM is found to make the void profiles shallower, with an inner density about 20\% larger than in the standard $\Lambda $CDM model, consistently with the previous results of \citet{Yang_etal_2015}, while for larger voids ($5$ Mpc$/h < R_{\rm eff} \leq 10$ Mpc$/h$) the impact of WDM is very mild, with the density profiles consistent at 2-$\sigma $ with the standard scenario. On the other hand, $f(R)$ gravity has the opposite effect of steepening the void profiles on both ranges of void size, even though the amplitude of the suppression of the inner void density is more pronounced for the smallest voids. As a result, the combined model featuring at the same time WDM and $f(R)$ gravity is strongly degenerate with $\Lambda $CDM for the smallest voids while resulting marginally distinguishable from the standard model for the larger ones. In this respect, we have extended the results of \citet{Yang_etal_2015} to the case of a non-standard theory of gravity, finding evidence of a strong degeneracy between the two models in the observables where the footprints of WDM first identified by \citet{Yang_etal_2015} are more pronounced.
\end{itemize}
\ \\

To summarise, we have performed and presented in this paper -- for the first time -- the results of a cosmological simulation of structure formation for a Warm Dark Matter particle candidate evolving under the effect of a modified theory of gravity, and compared its results to the reference $\Lambda $CDM cosmology and to the separate effects of these two modifications of the standard cosmology. 

We have performed an extensive analysis of the simulated matter distribution and of the properties of the resulting linear and non-linear structures, including halos, subhalos, and voids, and identified possible observational degeneracies between the models. We found that -- differently from the case of cosmologies featuring simultaneously modified gravity and massive neutrinos discussed in \citet{Baldi_etal_2014} -- most of the standard statistics do not show any significant degeneracy due to the fact that Warm Dark Matter and $f(R)$ gravity impact these statistics with very different functional dependencies on cosmic scales and halo masses. 

However, we also found that a clear degeneracy exists in the observational properties of small cosmic voids, with both their abundance and their density profiles being hardly distinguishable in the combined WDM-$f(R)$ simulation from the corresponding results in the standard $\Lambda $CDM cosmology, even if the two models show -- individually -- a clear and distinctive footprint in these observables.

\section*{Acknowledgments}
MB acknowledges support from the Italian Ministry for Education, University and Research (MIUR)
through the SIR individual grant SIMCODE, project number RBSI14P4IH.
The numerical simulations presented in this work have been performed 
and analysed on the Hydra cluster at the RZG supercomputing centre in Garching.

\bibliographystyle{mnras}
\bibliography{baldi_bibliography}

\label{lastpage}

\end{document}